\documentclass[%
 reprint,
superscriptaddress,
 amsmath,amssymb,
 aps,floatfix
]{revtex4-1}

\usepackage[pdftex]{graphicx}
\usepackage{dcolumn}
\usepackage{bm}
\usepackage{amsmath}
\usepackage{color}
\usepackage{float}
\usepackage{textcomp}

\begin{document}

\preprint{APS/123-QED}

\title{Flexible, photonic films of surfactant-functionalized cellulose nanocrystals for pressure and humidity sensing}

\author{Diogo V. Saraiva}
    \thanks{Authors contributed equally.}
    \affiliation{Department of Physics, Soft Condensed Matter and Biophysics, Debye Institute for Nanomaterials Science, Utrecht University, Utrecht 3584 CC, Netherlands}
\author{Steven N. Remi\"{e}ns}
    \thanks{Authors contributed equally.}
    \affiliation{Department of Physics, Soft Condensed Matter and Biophysics, Debye Institute for Nanomaterials Science, Utrecht University, Utrecht 3584 CC, Netherlands}
\author{Ethan I. L. Jull}
    \affiliation{Department of Physics, Soft Condensed Matter and Biophysics, Debye Institute for Nanomaterials Science, Utrecht University, Utrecht 3584 CC, Netherlands}
\author{Ivo R. Vermaire}
    \affiliation{Department of Physics, Soft Condensed Matter and Biophysics, Debye Institute for Nanomaterials Science, Utrecht University, Utrecht 3584 CC, Netherlands}
\author{Lisa Tran}
    \email{l.tran@uu.nl}
    \affiliation{Department of Physics, Soft Condensed Matter and Biophysics, Debye Institute for Nanomaterials Science, Utrecht University, Utrecht 3584 CC, Netherlands}

\begin{abstract}
Most paints contain pigments that absorb light and fade over time. A robust alternative can be found in nature, where structural coloration arises from the interference of light with submicron features. Plant-derived, cellulose nanocrystals (CNCs) mimic these features by self-assembling into a cholesteric liquid crystal that exhibits structural coloration when dried. While much research has been done on CNCs in aqueous solutions, less is known about transferring CNCs to apolar solvents that are widely employed in paints. This study uses a common surfactant in agricultural and industrial products to suspend CNCs in toluene that are then dried into structurally colored films. Surprisingly, a stable liquid crystal phase is formed within hours, even with concentrations of up to 50 wt.-\%. Evaporating the apolar CNC suspensions results in photonic films with peak wavelengths ranging from 660 to 920 nm. The resulting flexible films show increased mechanical strength, enabling a blue-shift into the visible spectrum with applied force. The films also act as humidity sensors, with increasing relative humidity yielding a red-shift. With the addition of a single surfactant, CNCs can be made compatible with existing production methods of industrial coatings, while improving the strength and responsiveness of structurally-colored films to external stimuli. 

\end{abstract}

\maketitle

\section{Introduction}
Mechanically strong and responsive coatings are important for extending the use and lifetime of materials, but for their future viability, the sustainability of the source material must be taken into account. Cellulose, being derivable from plant biomass, is one of the most widely available organic polymers \cite{Klemm2005,Tang2017,Li2021}. The crystalline regions of cellulose within the plant cell wall can be isolated via acid-hydrolysis, yielding chiral nanoparticles called cellulose nanocrystals (CNCs). At high concentrations, CNCs can self-assemble to form a helically-stacked, cholesteric liquid crystal that is capable of selectively reflecting light \cite{Revol1992}. The optical response of CNCs can be tuned by controlling the periodicity of their cholesteric assembly, characterized by the pitch $p$, which is the distance over which the CNC orientation rotates by 2$\pi$. The reflected light is circularly polarized and left-handed. The reflection wavelength $\lambda$ is related to the cholesteric pitch $p$ and can be estimated by assuming Bragg reflection from a perfectly ordered cholesteric \cite{deVries1951,Dreher1973}:
\begin{equation}
\lambda \approx n p \sin \theta,
\end{equation}
where $n$ is the average refractive index of the CNC and $\theta$ is the angle of light incidence. CNCs are thereby capable of a broad range of optical signals without needing to change solution components. The abundance and sustainability of CNCs and their ability to form photonic structures makes them attractive as replacements for synthetic polymers and pigments in coatings.

\begin{figure*}\centering
  \includegraphics[width=0.9\linewidth]{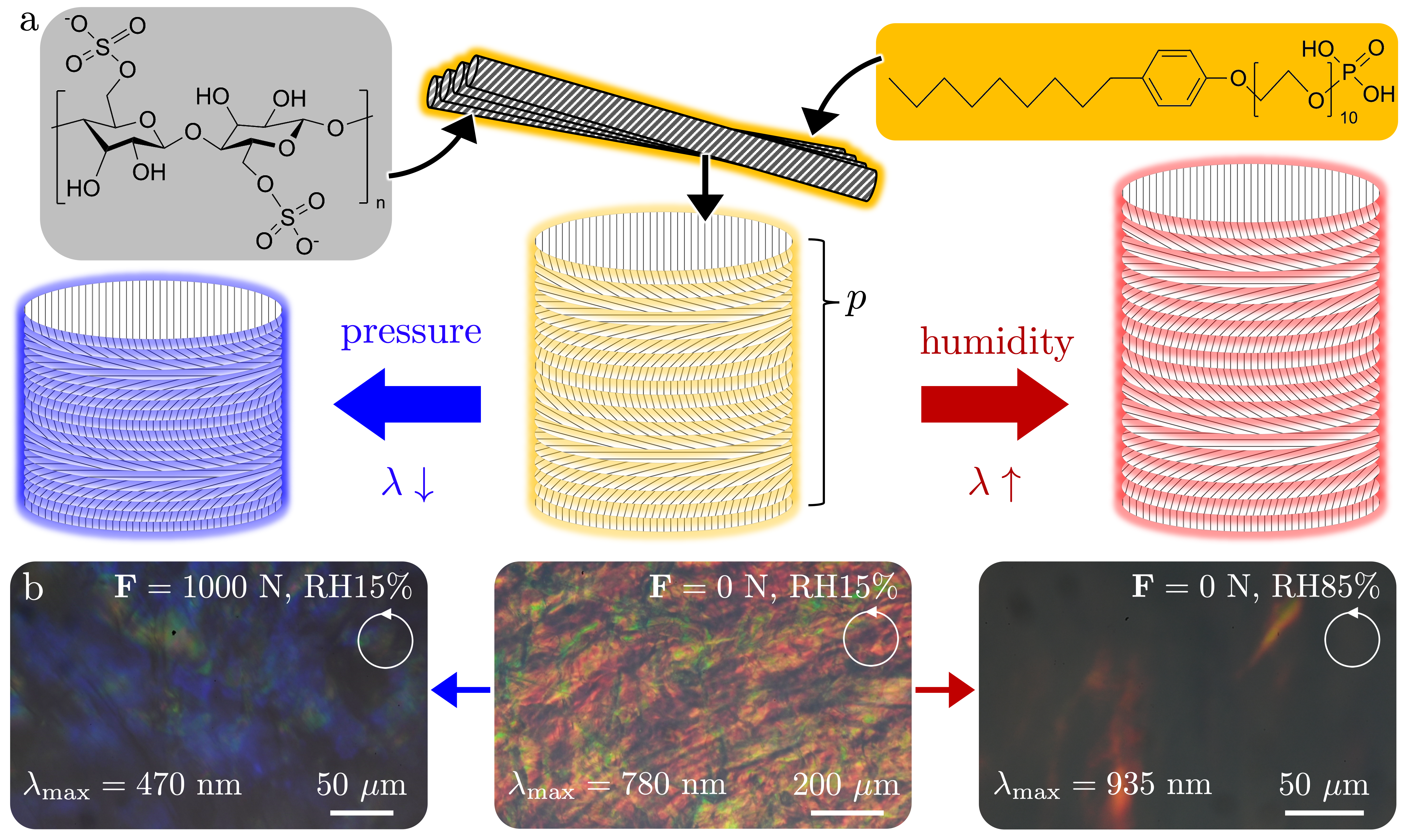}
  \caption{a) Schematics illustrating the self-assembly of surfactant-functionalized cellulose nanocrystals into cholesteric films. Crystalline regions of cellulose (top left), derived from cotton filter paper, can be isolated via sulfuric acid hydrolysis, imbuing the surface with sulfate half-ester groups. The isolated, chiral cellulose nanocrystals (CNCs, top center) are functionalized with a commercial surfactant STEPFAC™ 8170-U, allowing for stable dispersions in toluene. When the toluene is slowly evaporated, the CNCs are concentrated until they self-assemble into a chiral, cholesteric liquid crystal (bottom center). The helical ordering is characterized by the pitch $p$, which is the length for the CNC orientation to rotate by 2$\pi$. With applied force (bottom left) or higher relative humidity (bottom right), the pitch can be decreased or increased, respectively, giving rise to a corresponding blue- or red-shift in the reflected light wavelength. b) Micrographs of CNC films formed by the evaporation of toluene. The color of a reference film (center), measured via the peak wavelength $\lambda_\text{max}$ from UV-vis spectroscopy, can be blue-shifted with applied force, $\mathbf{F}$, (left) or red-shifted with increased relative humidity (RH) (right).}
  \label{fig1}
\end{figure*}

The majority of CNC studies have focused on their behavior in aqueous suspensions. CNCs are often isolated from biomass via sulfuric acid hydrolysis, resulting in several hydroxyl groups on the CNC surface being replaced by sulfate half ester groups (Figure 1a, top left) \cite{Tang2017,Ranby1951,Schutz2020}. The sulfate groups give rise to a strong, electrostatic, interparticle repulsion, enabling stable CNC suspensions in aqueous solutions. The suspensions can then be slowly dried over days to form a solid, photonic film (Figure 1). However, working with CNCs in aqueous solutions has several drawbacks. The assembly of CNCs in water is notoriously slow \cite{Delepierre2020}, potentially due to interparticle interactions that make them susceptible to gelation \cite{Bruckner2016} with kinetic arrest occurring at low concentrations of around 10 wt.-\% \cite{Schutz2020}. Films of neat CNCs cast from aqueous solutions are additionally brittle and fragile, making them vulnerable to damage and difficult to handle \cite{Schutz2020,Honorato-Rios2016,ATran2020}. Moreover, many industrial coatings require the use of apolar solvents, limiting the application of hydrophilic CNCs in these settings. 

To address these shortcomings, in this work, we examine the self-assembly of CNCs in apolar suspensions. The hydrophilic surface of CNCs can be functionalized with hydrophobic moieties to make the CNCs compatible with apolar solvents. Past studies have previously dispersed surface-modified CNCs in apolar solvents \cite{Hu2017,Wohlhauser2018,Delepierre2021,Mariano2014, Heux2000,Elazzouzi-Hafraoui2009,Frka-Petesic2017}, but only a few methods are capable of preserving cholesteric ordering when the suspension is concentrated. The most common surfactant for hydrophobizing CNCs that preserves cholesteric order is a commercial surfactant called Beycostat NA (BNA) \cite{Heux2000,Elazzouzi-Hafraoui2009,Frka-Petesic2017}. BNA is a surfactant that is a phosphoric ester of polyoxyethylene nonylphenyl ether, whose production has since been discontinued. Here, we functionalize sulfuric-acid hydrolyzed CNCs with an alternative, commercial surfactant that is also a phosphoric ester of polyethylene oxide (10) nonylphenyl, called STEPFAC™ 8170-U (Figure 1a, top right). STEPFAC-modified CNCs suspended in toluene can form a cholesteric phase within a matter of hours and at high concentrations of up to 50 wt.-\%. We then cast the apolar CNC dispersions to create photonic films, a system that is virtually unexplored to the best of our knowledge. The resultant films can be rapidly formed, are flexible, and exhibit distinct optical responses to applied force (Figure 1b, left) and changes in relative humidity (Figure 1b, right). The enhanced malleability and optical response of surfactant-modified CNC films to environmental stimuli has the potential for use in functional coatings, labels, and sensors.

\section{Results and Discussion}
\subsection{Surfactant-Functionalized CNCs in Toluene}

CNCs were obtained from the sulfuric acid hydrolysis of Whatman grade 1 cellulose filter paper. After the resultant aqueous suspension is dialyzed, the CNCs are prepared for transfer to an apolar solvent, namely toluene, following the procedure similar to the one first reported by Heux \textit{et al.}\cite{Heux2000}. Details are given in the Experimental Section and Supporting Information. In brief, the nonylphenol ethoxylated phosphate ester surfactant STEPFAC™ 8170-U (Figure 1a, right) is introduced to the aqueous CNC suspension in a 4:1 weight ratio (CNC:surfactant). The pH is adjusted to be between pH 8-9. The CNCs are then freeze-dried and resuspended in toluene. Excess surfactant is removed via centrifugation, and the supernatant is discarded. Fresh toluene is then introduced drop-wise until a qualitative reduction of the suspension viscosity is observed. The solution can be optionally tip-sonicated to disperse the STEPFAC-modified CNCs. 

To determine how dispersing the CNCs in toluene impacts the particle morphology, we analyze the size distribution of CNC particles before and after surfactant-functionalization using atomic force microscopy (AFM), shown in Figure 2a. The height distribution of the particles is plotted in Figure 2b, showing that the average radius of the CNCs is increased by the adsorbed surfactant. For neat CNCs, the average height is determined to be 10.4 $\pm$ 3.6  nm. On the other hand, for CNCs after functionalization, the average height is 12.7 $\pm$ 5.9 nm. By subtracting the average height of neat CNCs from that of STEPFAC-functionalized CNCs, we obtain an average surfactant layer of $\sim$1.2 nm. This measured surfactant thickness is comparable to that measured by Elazzouzi-Hafraoui \textit{et al.} using transmission electron microscopy \cite{Elazzouzi-Hafraoui2009} and Bonini \textit{et al.} using small-angle neutron scattering \cite{Bonini2002}, where BNA is used (Supporting Information) \cite{Heux2000,Elazzouzi-Hafraoui2009,Frka-Petesic2017,Bonini2002}.

To discern how the adsorbed surfactant affects the cholesteric assembly, we measure how the pitch varies with concentration by preparing a series of rectangular capillaries filled with the CNC-in-toluene suspension (Experimental Section). From a stock solution of 49 wt.-\%, we prepare dilutions with concentrations ranging from 18 to 47 wt.-\%. Interestingly, the resulting suspensions rapidly phase-separate into a cholesteric and an isotropic phase within an hour (Figure 2c, right). Pitch values of the cholesteric phase in each capillary are obtained from polarized optical micrographs (Figure 2c, insets). 

\begin{figure*}\centering
  \includegraphics[width=0.9\linewidth]{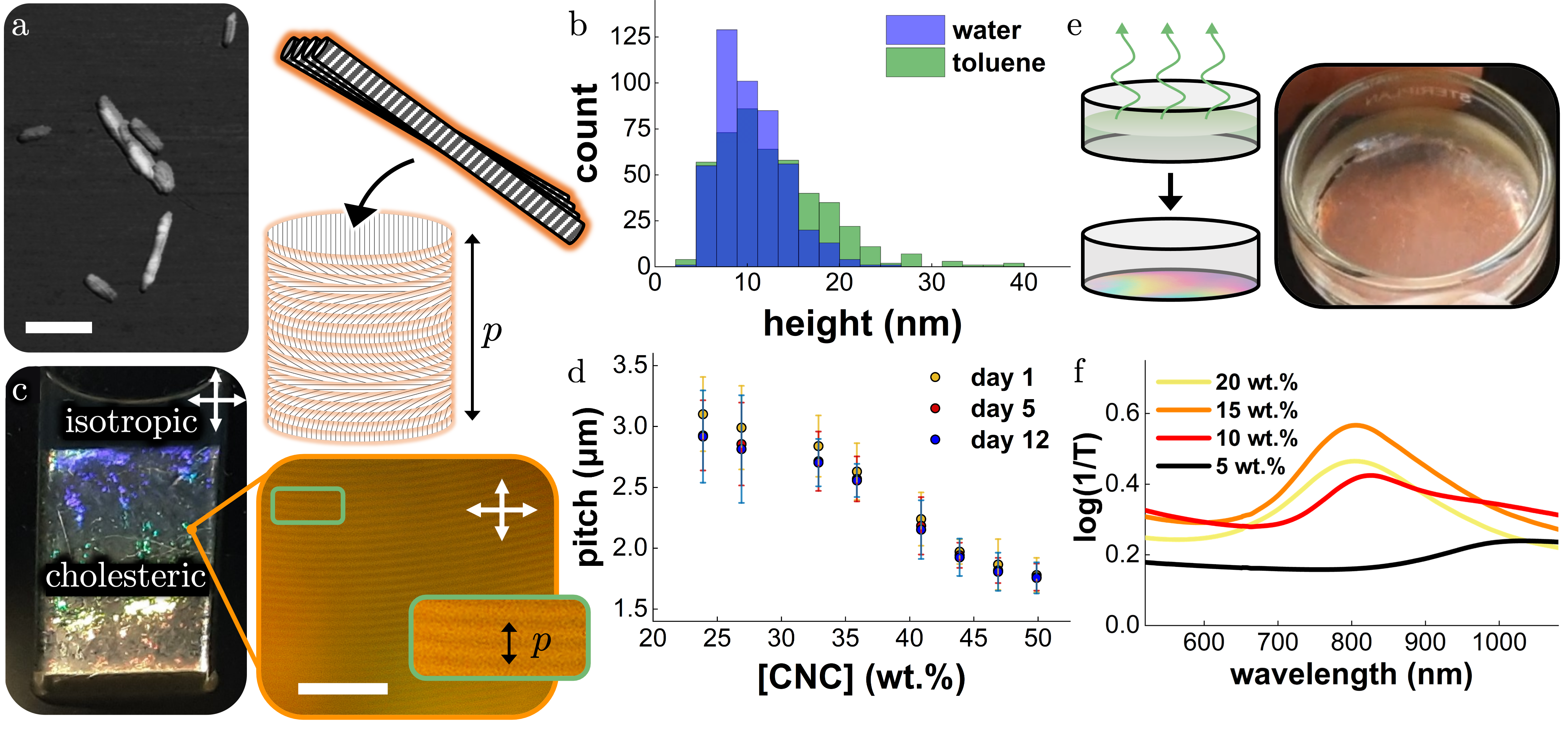}
  \caption{a) Atomic force microscopy (AFM) image of cellulose nanocrystals after functionalization with STEPFAC™ 8170-U. The phase signal is shown in the image, highlighting that bundles of cellulose nanocrystals are visible alongside individual crystallites. The scale bar is 200 nm. b) Histograms of AFM height distributions for CNC particles before and after functionalization with STEPFAC™ 8170-U. From the average difference before and after functionalization, the surfactant layer is determined to be approximately 1.2 nm thick. c) Polarized image of a capillary filled with a concentrated suspension of CNCs in toluene. In less than 24 hours, the suspension separates into a disordered isotropic (top) and an ordered cholesteric phase (bottom). The orange inset shows a polarized optical micrograph of the cholesteric phase, where stripes are visible due to the helical stacking of CNCs, depicted in the schematic above. The pitch size $p$ is labeled with a black arrow. The scale bar is 20 {\textmu}m. d) A plot of average pitch sizes with increasing CNC concentration after 1, 5, and 12 days. e) Schematic of the evaporation-assisted self-assembly of CNC films (left). A photograph of a CNC film formed via toluene evaporation is shown on the right. f) UV-vis spectroscopy data of films formed with varying initial CNC concentrations. The log of inverse transmittance (T) is plotted against light wavelength. With increasing concentration, the peak reflected wavelength begins to asymptote to $\sim 805$ nm.}
  \label{fig2}
\end{figure*}

The phase separation of the CNC-in-toluene suspension is unexpectedly fast compared to neat CNCs in water, which typically requires days to phase separate and weeks to fully equilibrate \cite{Honorato-Rios2016,Dong1996,Honorato-Rios2020}. Despite the increased thickness of each CNC particle, the average pitch value in toluene is significantly reduced compared to neat CNCs in water, in line with previous studies of CNCs in apolar solvents \cite{Heux2000,Elazzouzi-Hafraoui2009,Frka-Petesic2017,Bonini2002,Zhou2009}. The pitch values measured for CNCs in toluene at each concentration do not vary significantly after 1, 5, and 12 days, indicating that equilibrium pitch values can be reached within days, in contrast to the weeks required for neat CNCs in water (Figure 2d). A slight decrease in the pitch is measured on day 5 for lower concentrations. Changes in the cholesteric volume fraction for each concentration after 1, 5, and 12 days is also measured and corresponds to changes in the pitch values (SI Figure 1). The critical concentration to first obtain a cholesteric phase is 21 wt.-\%, and $\sim$40 wt.-\% is required for the suspension to become fully cholesteric. The increased fraction of the cholesteric phase between day 1 and day 5 is reflected in a jump in the biphase volume fraction (SI Figure 1). The biphase fraction does not change between 5 and 12 days, further supporting that the suspensions require only a few days to become equilibrated.

To determine the interaction energies between STEPFAC-CNCs in toluene, we measured the zeta potential and the conductivity of the suspension, in order to then estimate the Debye length (Supporting Information). For CNCs that have been washed three times via centrifugation, we measure a conductivity of 75.4 pS/cm (SI Table 1, SI Figure 2) and a zeta potential of around -10 mV (SI Figure 3). For an apolar solvent such as toluene with a low dielectric permittivity of $\epsilon_\text{r}\approx 2.38$, we expect the charged STEPFAC-CNCs to have a long-ranged electrostatic repulsion with a large Debye length. Assuming that the conductivity comes from excess hydrogen ions, we determine the ion concentration to be [H$^{+}$] = 1.4$\times$10$^{-7}$ (mol/m$^{3}$), yielding a Debye length of approximately 4.5 {\textmu}m. Note that because we have ignored bulkier, multivalent charged species as contributors to the measured conductivity, our Debye length approximation is likely an underestimation of the CNC's true Debye length in toluene. Yet, with the average length of STEPFAC-CNCs being 224.8 $\pm$ 113.7 nm (SI Figure 4), the approximated Debye length is still an order of magnitude larger than the particle itself. We note that past studies have attributed the colloidal stabilization of BNA-modified CNCs in apolar solvents to purely steric repulsion \cite{Heux2000,Elazzouzi-Hafraoui2009}. Here, we find from conductivity measurements that the electrostatic repulsion between CNCs that are functionalized with nonylphenol ethoxylated phosphate ester surfactants is significant and longer-ranged than purely steric repulsion. This remarkably long-ranged, Coulombic repulsion contributes to the prevention of STEPFAC-CNC gelation at high packing fractions.

Additionally, the large Debye length of our toluene system allows for low pitch values at high CNC weight percentages. The long-ranged, repulsive interactions of the STEPFAC-modified CNCs, combined with high packing fractions, allows for concentrated STEPFAC-CNC suspensions to be well-modeled by hard particle potentials. Low pitch values (under 2 {\textmu}m) with high mass fractions (45\% and higher) have been predicted from modeling CNCs as hard, chiral particles \cite{Chiappini2022,Sewring2023}, matching well the values measured in concentrated STEPFAC-CNCs in toluene (Figure 2d). On the other hand, much larger pitch values have been observed in aqueous systems (\textit{e.g.}, 7 {\textmu}m at around 7 wt.-\% \cite{Beck-Candanedo2005}).

Beyond aqueous systems, when comparing STEPFAC-CNCs in toluene to systems with other apolar solvents, the lowest pitch value in our toluene system ($\sim 1.5$ {\textmu}m) is lower than those previously reported \cite{Heux2000,Elazzouzi-Hafraoui2009,Frka-Petesic2017,Zhou2009}. The lower pitch values are due to two mechanisms. First, using STEPFAC™ 8170-U, we are able to disperse surface-functionalized CNCs in toluene with minimal tip sonication, a process known to increase the pitch \cite{Schutz2020,Frka-Petesic2023}. Second, toluene closely matches the refractive index of both the CNCs and the surfactant used. Toluene has a refractive index $n_\text{tol}\approx1.49$ \cite{Arif2017} as opposed to cyclohexane used in the work of Elazzouzi \textit{et. al.} \cite{Elazzouzi-Hafraoui2009} ($n_\text{cyc}\approx1.43$ \cite{Kerl1995}) or chloroform used in the work of Araki \textit{et. al.} ($n_\text{chl}\approx1.45$ \cite{Araki2001}). We measured the refractive index of the STEPFAC™ 8170-U surfactant to be $n_\text{STEP}\approx1.49$, using an Abbe refractometer (ATAGO NAR-3T). The closer index-matching of the solvent and the surfactant with crystalline, cotton-based cellulose ($n_\text{CNC}$ ranging from 1.53-1.60) \cite{Cranston2008} reduces van der Waals attraction \cite{Pusey1986} and facilitates more efficient packing of the CNC particles into a tighter helical structure. The reduced van der Waals attraction also contributes to the prevention of CNC gelation at high concentrations. Recent modeling of CNCs using classical density functional theory and hard rods further predicts possible pitch values that are below 2 {\textmu}m. In these models, an average CNC length of $\sim130$ nm is assumed \cite{Chiappini2022,Sewring2023}, which is within the length range of the CNCs used in this study (SI Figure 4). The STEPFAC-CNCs in toluene reported here are thus far the only system to approach these predictions, due to the reduced sonication needed for particle dispersion and the improved index matching of solution components.

\subsection{Photonic Films from Drying CNCs in Toluene}
To access the photonic properties of STEPFAC-CNCs, we formed films using evaporation-assisted self-assembly, depicted schematically in Figure 2e. Solvent evaporation of CNC suspensions leads to higher concentrations, reducing the pitch size (Figure 2d). Beyond a critical concentration, greater than 50 wt.-\% for the current system, the suspension gelates and kinetically arrests. Continued solvent evaporation leads to a compressive force that further reduces the pitch, resulting in submicron periodicities and a photonic response \cite{Honorato-Rios2016,Parker2016}. Films with varying initial concentrations, from 5 to 20 wt.-\% in steps of 5\%, are formed within covered glass petri dishes, $\sim$ 3 cm in diameter (Figure 2e). Covering the petri dish slows the evaporation from taking place within an hour to spanning over several days, allowing for reduced fluid flow and better equilibration of the cholesteric. Some films appear transparent at normal viewing angle but become uniformly red when tilted, indicating infrared light reflection with normal incidence (Equation 1). The reflectance properties of films with varying initial wt.-\% are measured using UV-vis spectroscopy (Figure 2f). With increasing initial concentration, the peak reflected wavelength is blue-shifted, asymptoting near 800 nm. The red-to-infrared, peak reflected wavelength of these films being larger than typical wavelengths of films formed from neat CNCs is due to the adsorbed surfactant, which increases the particle thickness by $\sim$2.3 nm, as well as the late onset of kinetic arrest, where higher concentrations are required for gelation to occur. Earlier kinetic arrest can facilitate a greater duration of pitch compression from evaporative stresses, which decreases the pitch size more dramatically than with increasing concentration \cite{Parker2016,Parker2022}. However, the point of kinetic arrest must be balanced against the time required for cholesteric equilibration. By varying the evaporation time from days to hours, we find a red-shift in STEPFAC-CNC films with shorter evaporation times (SI Figure 5).

The reflected light wavelength of the dried CNC film is directly dependent upon the pitch value of the CNCs in solution. To this end, we vary the pitch values in solution by washing via centrifugation, removing the supernatant, and resuspending the CNCs in fresh toluene (see Experimental Section and Supporting Information). Washing the CNC solution via centrifugation introduces additional sonication steps and removes excess ions, yielding larger Debye lengths --- both factors that increase the cholesteric pitch. The pitch values for two solutions of 20 wt.-\% STEPFAC-modified CNCs in toluene washed once and twice via centrifugation are determined using polarized optical micrographs. We measure the pitch values of the two solutions in capillaries and found that the pitch value for the solution washed twice increases compared to the one washed once, from 2.77 to 3.10 {\textmu}m. The increased pitch value translates to the increased reflected light wavelength when the solutions are dried into films. The films formed from the solution with one washing step has a peak reflected wavelength of 820 nm, while the solution with two washing steps has a peak reflected wavelength of 900 nm (SI Figure 6). We further note that by foregoing sonication in the surface-modification procedure entirely, CNC films with peak reflected wavelengths ranging from 650 to 750 nm are possible. Therefore, the reflected wavelength of a dried CNC film that is cast from an apolar suspension can be tuned by adjusting the solution conductivity and the degree of sonication. 

Notably, the resultant films are flexible and can be bent and slowly twisted without generating visible fractures (SI Video 1). This malleability is in stark contrast to films formed from neat, unfunctionalized CNCs, which are notoriously brittle due to strong hydrogen bonding between unfunctionalized CNCs \cite{ATran2020,Parker2018,Droguet2022,Andrew2023}. Past works have addressed the fragility of CNC films by co-assembly with additives in aqueous solutions, such as polyethelyne glycol (PEG) \cite{Bardet2015,Yao2017} and glycerol \cite{He2018}. For these and similar systems, the films can have an optical response to environmental changes, but the concentration of the additive must be carefully tuned to preserve the cholesteric ordering. In a system less sensitive to additive concentration, another work has functionalized CNCs with zwitterionic surfactants to increase the film flexibility, but no sensitivity to external stimuli is reported \cite{Guidetti2016}. The surfactant used in this work, necessary to disperse the CNCs in apolar solvents, both alters the mechanical properties of the assembled film and enables an optical response to compression and relative humidity --- all from a single surface-modification process. 

\subsection{Pressure Response}

We determine the response of STEPFAC-CNC films to applied forces by compressing the films with a manual hydraulic press and measuring the photonic response with reflectance spectroscopy (Experimental Methods and Supporting Information). For low applied forces, the films are compressed using a commercial, hand-cranked, force testing stand with an integrated, digital force reader. For forces above 500 N, a hydraulic press is used. A button load cell is placed between the piston of the hydraulic press and the sample to quantitatively measure the compressive force (Figure 3a, center; see also SI Figure 7). Figure 3b shows images of a film made from an initial STEPFAC-CNC concentration of 20 wt.-\% in toluene after sequential compression steps. The peak reflected wavelength is reported at the top of each image. Reflectance spectroscopy measurements after each compression step for three different films are plotted in Figure 3c, showing that applied force blue-shifts the reflected wavelength. We additionally image cross sections of the film before and after complete compression using scanning electron microscopy (SEM), shown in Figure 3d.

\begin{figure*}\centering
  \includegraphics[width=0.95\linewidth]{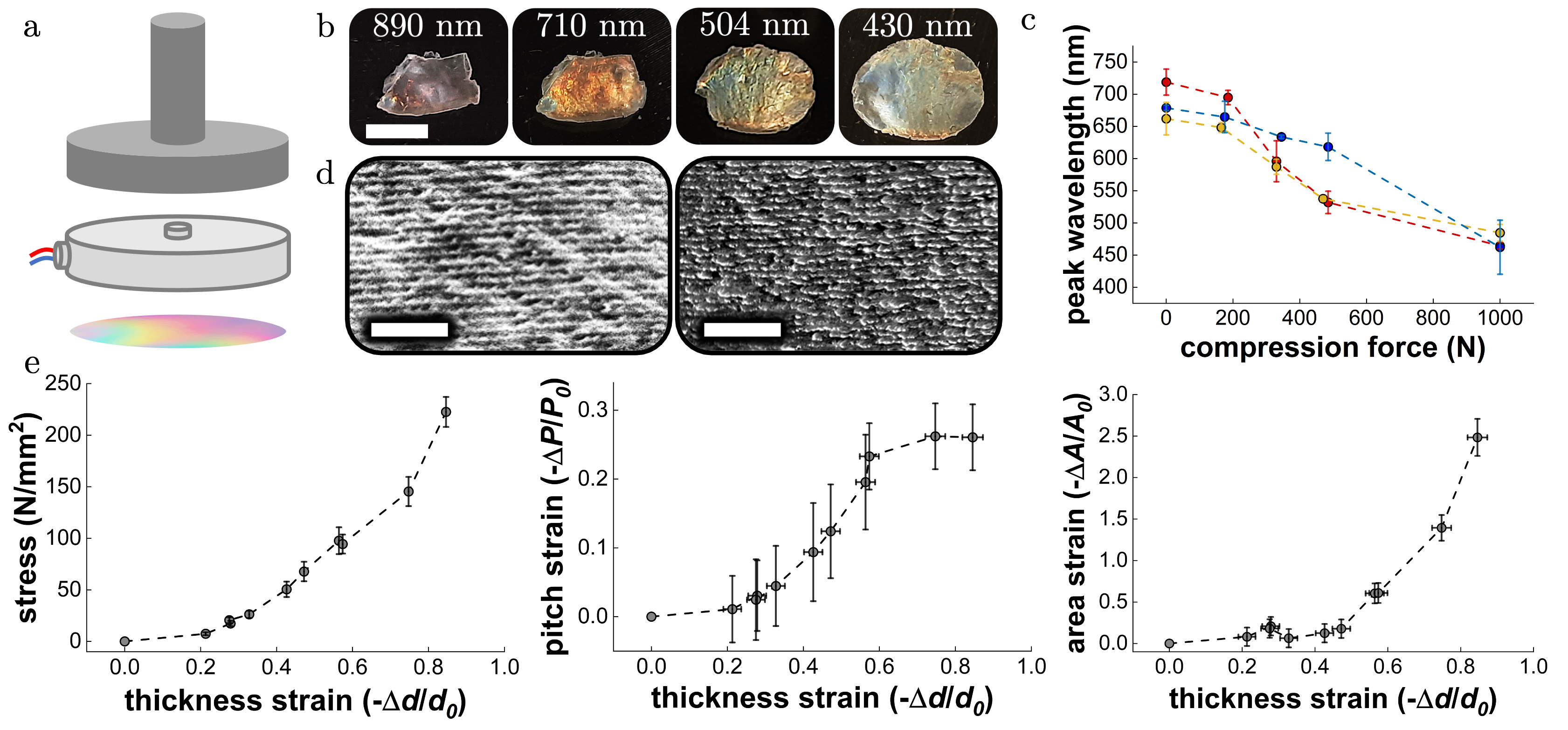}
  \caption{a) Schematic of applied force experiments for forces above 500 N. The CNC film formed from a 20 wt.-\% toluene suspension is placed within a hydraulic press with a button force sensor. b) Photographs of films with increasing applied force from left to right. The scale bar is 4 mm. The peak reflected wavelength of each film is reported at the top. c) The peak wavelength as measured by reflectance spectroscopy data of CNC films undergoing compression is plotted against sequential steps of applied force, as quantified by the force sensor. The data from three different films are plotted, colored red, yellow, and blue. With increasing steps of applied force, the reflected light blue-shifts, with possible wavelengths spanning the visible range. d) Scanning electron micrographs of film cross sections before (left) and after (right) the series of compressions. Before compression, the pitch size is measured to be $\sim500$ nm ($525 \pm 24$ nm) and has a reflected wavelength of $\sim 805$ nm. After compression, the pitch size is measured to be $\sim390$ nm ($385 \pm 35$ nm) and has a blue-shifted reflected wavelength of $\sim710$ nm. The scale bar is 2 {\textmu}m. e) Stress-strain curves obtained from film compression.}
  \label{fig3}
\end{figure*}

Interestingly, the deformations induced by compressive forces on STEPFAC-CNC films remain even after the force is removed, indicating that the film supports plastic deformations. We demonstrate how the film retains deformations from compression by stamping a STEPFAC-CNC film (SI Figure 8). The stamped region is noticeably blue-shifted compared to the red, uncompressed film, as reflected in the UV-vis spectroscopy measurements of the film, plotted in SI Figure 8. Furthermore, the degree of sonication in the surface-functionalization procedure impacts the mechanical strength of the final film. For films with starting wavelengths in the infrared (Figure 3b), the STEPFAC-CNCs undergo sonication, resulting in a visible optical response only for high forces of around 1500 N (SI Figure 11). For STEPFAC-CNCs dispersed in toluene without sonication, only forces on the order of around 100 N are needed for a visible optical response. The mechanical properties of STEPFAC-CNC films are variable, and the precise tuning of the mechanical properties with the surface-functionalization procedure is the subject of a future study. 

We characterize the material properties of unsonicated STEPFAC-CNC films using the force testing stand, which allows for the precise application of forces up to $\sim 500$ N. We determine the stress-strain curve from sequential compression steps by measuring the applied force, the area, and the thickness of the film after each step. In Figure 3e, the applied stress versus the thickness strain is plotted on the left, where $\Delta d$ is the change in thickness, and $d_0$ is the thickness of the uncompressed film. After the first compression, the film undergoes plastic deformations, shown by the flat region of the curve. The stress-strain curve does not exhibit a linear elasticity regime. With increasing stress, the plateau of the plastic deformation ends. The strain begins to increase sharply, indicating densification of the sample, as reflected by a reduction of the film's volume (SI Figure 9). From reflectance spectroscopy measurements taken after each compression step and Equation 1, we determine the strain in the pitch and plot it against the thickness strain, where $\Delta p$ is the change in the pitch value and $p_0$ is the pitch of the uncompressed film (Figure 3e, center). Interestingly, the pitch strain reduction only occurs in the range of thickness strains where the film undergoes densification. At higher thickness strains, where the pitch strain begins to plateau, we see significant increases in the film's area, shown in the area versus thickness strain graph on the right of Figure 3e, where $\Delta A$ is the change in area and $A_0$ is the area of the uncompressed film. The increased strain with more compression causes the film to spread and increase its area, rather than to change the pitch. Therefore, the optical response of the film and the change in pitch from compression are a direct result of densification of the STEPFAC-CNC film with applied force.

The shift in the reflected wavelength with compression is directly proportional to the pitch value, according to Equation 1. We directly observe how the pitch of the STEPFAC-CNC film is altered with densification by imaging film cross sections using SEM (Figure 3d). The film is flash-frozen with liquid nitrogen to rigidify it before fracturing by hand. The film is then sputter-coated with a 5 nm layer of platinum to reduce charging and improve image contrast. We extract a pitch size from SEM images by performing 30 manual measurements using ImageJ (Experimental Section). The pitch size for an uncompressed film formed at an initial concentration of 20 wt.-\% (Figure 3d, left) is determined to be $525 \pm 24$ nm. With the film having an average refractive index ranging from 1.5-1.6 \cite{Cranston2008} and at normal incidence, a reflected wavelength range of 752-878 nm is expected (Equation 1). The peak reflected wavelength of 805 nm, as measured by UV-vis spectroscopy (SI Figure 10), falls within this range. For a film compressed for 1 minute with 1500 N of force, we measure a pitch of $385 \pm 35$ nm from SEM images. A reflected wavelength between 525 and 672 nm is expected, but the peak reflected wavelength as measured by UV-vis spectroscopy is 710 nm. That the measured reflected wavelength falls outside of the predicted range is unsurprising. The film thickness is not uniform across the film, resulting in uneven film compression. The variation in the amount of compression exerted across the film is evident in the multiple colors seen within the right two images of Figure 3b and the broadening of reflected light wavelengths with increasing compression in SI Figure 11. In general, due to densification and pitch reduction of the assembled STEPFAC-CNCs with compression, the reflected wavelength of the film is modifiable with applied pressure. 

\begin{figure*}\centering
  \includegraphics[width=0.9\linewidth]{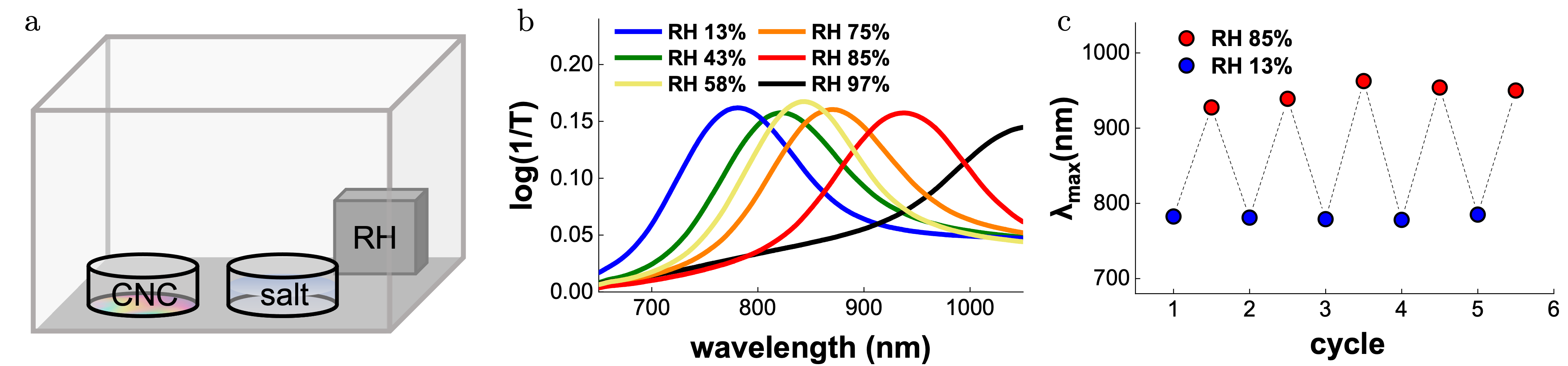}
  \caption{a) Schematic of a humidity chamber, where the CNC film is placed in a box with a saturated salt solution and a digital hygrometer. b) UV-vis spectroscopy data of an initially 20 wt.-\% STEPFAC-CNC film, left to equilibrate in the humidity chamber for varying relative humidity (RH) values, as set by the saturated salt solution (SI Table 3). The log of inverse transmittance (T) is plotted against light wavelength. With increasing RH values, the reflect light red-shifts. c) Plot of the peak reflected wavelength for sequential RH cycles between 13\% and 85\%. For each RH value, the CNC film returns to the same reflected wavelength range for at least 5 cycles, demonstrating a robust optical response to humidity.}
  \label{fig4}
\end{figure*}

\subsection{Humidity Response}

The STEPFAC-functionalized films are not only optically responsive to pressure but also to moisture in their environment. Figure 4a shows a schematic representation of a chamber used to test the film's optical response to varying relative humidity (RH). A solution of saturated salt is used to fix the RH value within the chamber, which is a typical method employed to calibrate commercial hygrometers. Each salt, the amount used, and the corresponding RH value obtained in the chamber are given in SI Table 3. A digital hygrometer is additionally included in the chamber to confirm and record the RH value. After the film is left to equilibrate in the chamber, the film is then removed and immediately placed within the UV-Vis spectrometer to measure the reflected light wavelength, plotted in Figure 4b. By recording the reflected light wavelength of a film over time, the optical response of the film equilibrates after approximately 2 hours (SI Figure 12). 

For a film formed from initially 20 wt.-\% STEPFAC-CNCs in toluene exposed to 13-97\% RH, the wavelength shifts from 781 to 1056 nm, giving a $\sim275$ nm optical change from dry to moist environments. To assess how robust the STEPFAC-CNC film's optical response is to humidity, we expose it to repeated cycles of dry (13\% RH) and moist (85\% RH) environments and record the peak reflectance from UV-vis spectroscopy after each exposure (Figure 4c). The film returns to the same range of reflected wavelengths for at least 5 cycles. For dryer conditions, the reflected wavelength only deviates by 7 nm around 781 nm, while the optical response to 85\% RH typically falls within a 30 nm window around 945 nm. STEPFAC-CNC films give a robust and distinct optical response to changes in the relative humidity. 

The STEPFAC surfactant (Figure 1a, top right) amplifies the assembled film's sensitivity to humidity compared to others made from aqueous suspensions. Neat, unfunctionalized CNCs are already hygroscopic from having hydrophilic, hydroxyl groups on their surface. However, films from neat CNCs only have a narrow red-shift of 100 nm from dry to wet \cite{Zhang2013}. Those formed with polyacrylamide \cite{Lu2017} improve the water absorption of the film slightly, increasing the red-shift amount to approximately 150 nm. Notably, a past work by Yao \textit{et al.} reports a humidity-responsive CNC film with 20 wt.-\% PEG ($M_\textrm{w} = 20$ kDa), where the optical response to humidity was enhanced by over a 250 nm range, from green for 30\% RH to red for 95\% RH \cite{Yao2017}. The STEPFAC surfactant used in the current work has a polyethylene oxide (10) block, which is molecularly similar to PEG. The STEPFAC surfactant is equally as hygroscopic as PEG, corroborated by the similar degree of swelling seen in the system of Yao \textit{et al.} with high RH. From weighing a STEPFAC-CNC film in different relative humidity conditions, we further confirm that the weight of the film increases with the relative humidity due to water absorption (SI Figure 13). The current STEPFAC-CNC system additionally has the advantage of being compatible with the existing infrastructure for most industrial coatings, where hydrophobic, polymeric matrices are commonly employed.

The optical response of a STEPFAC-CNC film to humidity can be changed by compressing the film, as shown in SI Figure 14. A film formed from 20 wt.-\% STEPFAC-CNCs is compressed with a force of 500 N for 5 seconds, blue-shifting the peak reflected wavelength of the film from $\sim800$ nm to around $\sim500$ nm. From exposure to repeat cycles of 13\% to 85\% relative humidity, the compressed film exhibited a significantly reduced peak reflected wavelength range of $\sim560$ nm for 13\% relative humidity and $\sim590$ nm for 85\% relative humidity. Accordingly, the sensitivity of STEPFAC-CNC films to humidity is adjustable, demonstrating the material's versatility for varying applications.

\section{Conclusion}

To conclude, CNCs can be made hydrophobic using a commercial nonylphenol polyethylene oxide phosphate ester surfactant, STEPFAC™ 8170-U, making the CNCs dispersible in apolar solvents and increasing their compatibility with the industrial production methods of coatings. Using toluene as the solvent to better index match the solution components, we obtain cholesteric pitch values of STEPFAC-CNCs as low as $\sim1.5$ {\textmu}m in solution. With evaporation-assisted self-assembly, we produce structurally-colored, flexible, and malleable films that are photonically responsive to applied force and humidity. The mechanical strength of the film can be selected by tuning the surface-functionalization procedure. Moreover, the optical response of the film to relative humidity can be modified through compression of the film. We further report, for the first time, an extraordinarily large Debye length for CNCs in apolar solvents. The long-ranged, electrostatic repulsion prevents gelation at high concentrations and facilitates fast phase separation kinetics, both characteristics that are advantageous for manufacturing CNC-based materials with photonic properties. 

Our STEPFAC-CNC system has several attributes that are desirable for a broad range of applications. We expect that the fast phase separation kinetics and apolar character of our STEPFAC-CNC system are valuable properties for roll-to-roll processes \cite{Droguet2022,Sondergaard2013}. Moreover, the microfluidic production of CNC-based materials further benefits from a reduction in the amount of organic solvent needed for flow focusing and the fast assembly kinetics seen in our system \cite{Parker2016,Parker2022,Suzuki2018,Hausmann2020}. Encapsulating STEPFAC-CNCs into emulsions can additionally increase the system surface area, further enhancing sensitivity to relative humidity. Lastly, STEPFAC-CNCs, being suspended in toluene, have the potential for their assemblies to be manipulated with applied electric fields, which can further regulate the assembly structure towards the specific needs of an application \cite{Frka-Petesic2017,Frka-Petesic2023}. We anticipate that the myriad beneficial properties of STEPFAC-functionalized CNCs will broaden the capabilities and use of biodegradable CNCs in industrial and commercial settings.

\section{Experimental Section}

\subsubsection{CNCs Sourced from Cotton Filter Paper}
Cellulose nanocrystals were obtained from the acid hydrolysis of Whatman no.1 cellulose filter paper (35 g) with 490 mL of 64 wt.-\% sulfuric acid (Sigma-Aldrich, 95-98\%) at 66 °C. The cellulose source was shredded beforehand with a spice grinder (Waring WSG60K). The acid hydrolysis reaction was quenched after 30 minutes with approximately 5 L of deionized water. The resulting suspension was allowed to sediment overnight, and the clear supernatant was decanted. The sedimented material was centrifuged in cycles of 20 minutes at 15000 RCF until all of it was in pellet form. Subsequent centrifugation cycles consisted in adding DI water until the CNCs in the supernatant suspended. Only the supernatant was collected and the pellet was discarded. This suspension was dialyzed over three weeks in DI water (replenished daily) with MWCO 12–14 kDa dialysis membranes (Spectrum™ Spectra/Por™ 4). This procedure yielded a 1.13 wt.-\% CNC suspension with a total volume of 900 mL. 

\subsubsection{Suspension of CNCs in Toluene}
STEPFAC™ 8170-U surfactant (Stepan) was added to the aqueous CNC suspension in a proportion of 4:1 (w/w) of surfactant to CNC, and the suspension was left stirring overnight. The suspension's pH was then adjusted using a 1 wt.-\% NaOH solution, until the measured pH was around 8-9. The $\sim$1.2 L suspension was then sonicated for 10 minutes at 44 W (Hielscher UP200St, 22 J/mL total sonication dose). The suspension was freeze-dried in 200 mL steps, each over 48 hours. The resulting dried CNC pellets were suspended in toluene (Thermo Scientific, 99.5\%\ ACS reagent) at a 2.9 wt.-\% concentration. Excess surfactant was removed by centrifugation at 15000 RCF over 90 minute cycles. With each cycle, the supernatant was discarded. After all of the CNCs were in pellet form, a small volume of toluene was carefully added dropwise until the pellet was resuspended. Some of the resulting stock suspension ({[CNC]=49.9 wt.-\%}) was tip-sonicated for 22 J/mL of suspension (33 W) in three 4-second steps, in an ice bath, to prevent sample overheating. 

\subsubsection{CNC Phase Diagram and Pitch Size Measurements} From the stock suspension, 200 {\textmu}L samples were prepared in a series of concentrations (50, 47, 44, 41, 36, 33, 31, 29, 27, 24, 21, and 18 wt.-\%). Each sample was enclosed in a rectangular glass capillary (1x10x50 mm, Vitrocom), sealed using epoxy glue (Pattex Super Mix Metal). Capillary photographs were taken with the sample placed between two crossed polarizers, and ImageJ software was used to measure the isotropic-cholesteric biphase fraction. Cross-polarized micrographs were captured with a Nikon DS-Ri2 CMOS camera attached to a Nikon Ti-E inverted microscope, equipped with a T-P2 DIC Polarizer module and an analysis polarizer. Pitch sizes for the cholesteric phase of each sample were determined from 30 manual measurements (in ImageJ) across three cross-polarized micrographs. Only the lowest region of the cholesteric phase (height-wise) was imaged. Biphase fractions and pitch sizes were measured after 1, 5, and 12 days of sedimentation. 

\subsubsection{CNC Film Formation} Diluted samples of the stock surfactant-coated CNC suspension were prepared at different concentrations (5, 10, 15, 20 wt.-\%). Each CNC film was produced by evaporating 2 mL of these suspensions in a lid-covered glass petri dish (40x12 mm, Duran Steriplan). The evaporation process typically required 4 to 5 days for the sample to be completely dry. 

\subsubsection{Conductivity Measurements} A Model 627 conductivity meter (Scientifica) was used for all conductivity measurements by filling the receptacle with 3 mL of sample. The CNC suspension samples had a concentration of 3 wt.-\%, and the concentration of the supernatant samples was as collected from the centrifugation step.  

\subsubsection{Reflectance and Transmission Spectroscopy} Reflectance spectra for the films were obtained using an OceanOptics OceanHDX miniature spectrometer mounted onto a Leica DM IL LED Inverted Microscope. A wavelength range of 420 to 800 nm was analyzed. UV-Vis transmission spectra for the films were obtained using a LAMBDA 365 UV/Vis Spectrophotometer. For all spectra, the rate of acquisition was 960 nm/min, in a wavelength range of 400 to 1100 nm. A rectangular region of about 1 cm by 0.5 cm was analyzed. 

\subsubsection{Optical Microscopy of Films} The colored films were imaged in reflection on a Leica DM IL LED inverted microscope with a left circular polarizer in the light path. Pictures were taken using a mounted Nikon Z6 digital camera. 

\subsubsection{Atomic Force Microscopy of CNCs} The size distribution of CNC suspensions was obtained using atomic force microscopy (AFM, JPK Nanowizard 2 in tapping mode), equipped with silicon AFM probes (OTESPA-R4, Bruker, scanning frequency 300 kHz, k=26 N/m). From the AFM images, the sizes of 466 particles were measured using the software Gwyddion (version 2.62, http://gwyddion.net/). The AFM imaging samples were prepared by dropcasting 200 {\textmu}L of CNCs in water and CNCs in toluene (both at 0.004 wt.-\%) onto freshly cleaved mica (Ted Pella, V1 Mica, 25 x 25mm). See Supporting Information for additional details on sample preparation and particle size analysis.  

\subsubsection{SEM Imaging} The morphological properties of CNC films were observed using a TFS Helios Nanolab G3 scanning electron microscope (SEM) in high vacuum mode at 2 kV, with a working distance of 5-5.5 mm. To observe the layered structure of the CNC films, small regions from uncompressed and compressed films were collected, submerged in liquid nitrogen for one minute, and cracked open with two tweezers to expose the cross-section. For SEM imaging, the samples were attached to conductive carbon tape and sputter-coated with a 5 nm layer of platinum. 

\subsubsection{Quantifying Applied-Force Response of Films} Force application on the CNC films up to 500 N was performed with a hand-cranked force test stand (Baoshishan) with an attached digital force gauge (Baoshishan 500 N Force Meter). At forces above 500 N, a manual hydraulic press (Specac, 15 ton) was used in conjunction with a button load cell (Phidgets, 0-1000 kg) to externally measure applied force. In all trials, small pieces of CNC films were compressed between strips of teflon tape, to prevent the sample from sticking to other surfaces. Before each trial, the samples were dried in an oven at 40 $^\circ$C for at least 20 minutes (at relative humidity values between 10\%\ and 15\%, as measured by a commercial hygrometer). 

\subsubsection{Mechanical Response Analysis} During applied-force trials, the film's thickness, projected area, and reflectance wavelength were monitored after each compression step. These trials were performed on individual pieces of solid film, around 0.7 to 1 cm in diameter, which were extracted from the petri dish using a scalpel. The film's thickness was measured using a micrometer. The film's projected area was measured from pictures using a commercial phone camera (Samsung Galaxy S21 FE). To obtain an accurate measurement of the film's area, color thresholding was applied to each photograph using ImageJ software. This thresholding automatically sets a mask around the perimeter of the film, and the film's area can be estimated. The pitch size of the film was estimated using Equation 1, where $n=1.55$, $\theta$ $=$ 90$^\circ$, and $\lambda$ equals the average peak reflected wavelength, as obtained from reflectance spectroscopy. These peak wavelength values were obtained by averaging five reflectance spectra, measured with a OceanOptics OceanHDX miniature spectrometer mounted onto a Leica DM IL LED inverted microscope.

\subsubsection{Quantifying Humidity Response of Films} CNC films were exposed to specific relative humidity (RH) values (from 13\% to 97\%) by sealing them (for a minimum of 6 hours) inside a small plastic box along with a petri dish filled with one of several saturated salt solutions (listed in SI Table 3). 2\% relative humidity was obtained by sealing the films in a desiccator with fresh silica beads.

\begin{acknowledgments}
We thank Stepan Company for providing a sample of STEPFAC™ 8170-U. We thank Kelly Brouwer, Ali Kosari, Mies van Steenbergen, Relinde van Dijk-Moes, and Dave van den Heuvel for experimental support. We thank Alfons van Blaaderen, Maarten Bransen, Marjolein Dijkstra, Arnout Imhof, Allard Mosk, and Tor Sewring for useful discussions. D. V. S. and I. R. V. acknowledge financial support from the Department of Physics, Utrecht University. E. I. L. J. acknowledges funding from the European Commission (Horizon-MSCA, Grant No. 101065631). L. T. acknowledges support from the European Commission (Horizon-MSCA, Grant No. 892354) and the NWO ENW Veni grant (Project No. VI.Veni.212.028). D. V. S. and L. T. acknowledge support from the Starting PI Fund for Electron Microscopy Access from Utrecht University's Electron Microscopy Center.
\end{acknowledgments}

%

\end{document}